\documentclass{article}
\begin{document}

 \title{Nonadiabatic Factor Accompanying Magnetic Translation of a Charged Particle}

 \author{J. Chee\\\\Department of Physics, Tianjin Polytechnic University,
\\Tianjin, 300160, China}

 \date{June 28, 2007}
 \maketitle
 \begin{abstract}
 The quantum adiabatic theorem incorporating the Berry phase phenomenon
 can be characterized as a factorization of the time evolution operator into
 the product of a path-dependent geometric factor, a usual
 dynamical factor and a nonadiabatic factor that approaches the
 identity in the adiabatic limit. We study a case where all these
 factors can be constructed explicitly and where the instantaneous
 Hamiltonian has infinitely degenerate eigenstates associated with magnetic translation symmetry.
 \end{abstract}
 \pagebreak


Magnetic translation symmetry is much discussed in the
literature\cite{brown1,zak1,zak2,zak3,geyler,exner,chee}. The basic
example is the quantum mechanics of a charged particle moving in a
two-dimensional plane perpendicular to a magnetic field, i.e. the
Landau level problem. The Hamiltonian is not invariant under the
ordinary translation group $R^2$, because it is the vector potential
rather than the magnetic field that appears in the Hamiltonian. The
translational symmetry of the physical situation is realized through
the magnetic translation. The magnetic translation concept has been
found useful in a variety of physical situations including Bloch
electrons in a magnetic field\cite{zak4,brown2,kohmoto,goryo} where
there is a lattice potential present. Even so, we believe that
magnetic translation symmetry as an essential tool in solving simple
dynamical systems have yet to be fully explored.

In this paper, we want to show a connection between magnetic
translation symmetry and the quantum adiabatic theorem involving
infinitely degenerate eigenstates. We adopt the point of view that
the quantum adiabatic theorem including the Berry phase phenomenon
is essentially a factorization of the time evolution operator into
the product of a path-dependent geometric factor and a usual
dynamical factor\cite{messiah, chee}. When the change of the
parameters in the Hamiltonian is not slow, there should be another
factor in the time-evolution operator which represents nonadiabatic
effects. Naturally, this nonadiabatic factor approaches the identity
operator in a suitable adiabatic limit. We want to provide an
example where all these factors can be constructed explicitly for a
general parameter variation and where the instantaneous Hamiltonian
has infinitely degenerate energy levels.

First, we collect some basic properties on the magnetic translation
to be used later. In the usual two-dimensional Landau level problem,
the kinematical momentum is $\pi_\mu=p_\mu-\frac{q}{c}A_\mu({\bf
x}),\ \mu=1, 2$, where $A_\mu(\bf x)$ is the vector potential in
arbitary gauge. Define
\begin{equation}
\eta_{1}=\pi_1-\frac{qB}{c}x_2=-\frac{qB}{c}c_2, \ \
\eta_{2}=\pi_2+\frac{qB}{c}x_1=\frac{qB}{c}c_1.
\end{equation}
In classical mechanics, $(c_1, c_2)$ is the center of the circular
motion. In quantum mechanics, we have the following commutation
relations
\begin{equation}
[\pi_1, \pi_2]=i\hbar qB/c,\ \ \ [\eta_{1}, \eta_{2}]= -i\hbar
qB/c,\ \ \ [\pi_\mu, \eta_{\mu}]=0.
\end{equation}
To realize a translation ${\bf x}\rightarrow {\bf x}+{\bf d}(t)$,
instead of using the ordinary translation operator $\exp(-ip_\mu
d_{\mu}(t)/\hbar)$ which does not commute with $\pi_\mu$ and the
Hamiltonian, one can choose to use the magnetic translation operator
$\exp(-i\eta_{\mu} d_{\mu}(t)/\hbar)$. Note that a distinction
between $\exp(-i\eta_{\mu} d_{\mu}(t)/\hbar)$ and
$P\exp(-i\eta_{\mu} d_{\mu}(t)/\hbar)$ has to be made because
$\eta_1$ and $\eta_2$ do not commute. Let
\begin{equation}
P\exp(-i\eta_{\mu} d_{\mu}(t)/\hbar)=e^{i\beta(C(\bf
d))}\exp(-i\eta_{\mu} d_{\mu}(t)/\hbar).
\end{equation}
Then the phase $\beta(C(\bf d))$ is determined by the path $C$
traversed by ${\bf d}(t)$. In particular, for a closed path,
$\beta(C(\bf d))$ is equal to $-\frac{q\phi}{\hbar c}$, where $\phi$
is the magnetic flux enclosed by the loop of $C$. One can similarly
consider path-ordered or time-ordered operators generated by
$\pi_\mu$. The magnetic translation group is defined by Zak as the
set of path-ordered magnetic translation operators\cite{zak1,zak2}.

Now let us consider a charged particle moving in a two-dimensional
plane with ${\bf x}=(x_1, x_2)$ whose Hamiltonian depends on a
parameter ${\bf R}(t)=(R_1(t), R_2(t))$:
\begin{equation}
H_L=\frac{1}{2m}\big[{\bf p}-\frac{q}{c}{\bf A}_L(x, {\bf
R})\big]^2,
\end{equation}
where
\begin{equation}
{\bf A}_L(x, {\bf R})=\frac{B}{2}{\bf e}_3\times\big[{\bf x}-{\bf
R}(t)\big].
\end{equation}
The Hamiltonian $H_L$ for a fixed ${\bf R}$ is just a Landau level
Hamiltonian with the magnetic field direction ${\bf e}_3$
perpendicular to the $(x_1, x_2)$ plane. Our purpose is to study the
case when ${\bf R}(t)$ varies with time.

Now a crude estimation of how the wave packet moves. The vector
potential, in addition to the uniform magnetic field in the ${\bf
e}_3$ direction, represents an electric field ${\bf E}=
-\frac{1}{c}\frac{\partial {\bf A}}{\partial t}= \frac{B}{2c}{{\bf
e}_3}\times \dot{\bf R}(t).$ To balance out this electric field, the
charged particle may generate a Lorentz force with a velocity
$\dot{\bf R}(t)/2$. Therefore it makes sense to conjecture that the
wave packet is shifted by $({\bf R}(t)-{\bf R}(0))/2$ at time $t$
instead of ${\bf R}(t)-{\bf R}(0)$. This picture should be correct
at least for the special case of a constant electric field which is
a well understood example in quantum mechanics. There has also been
study of the special case when the electric field is along a fixed
direction and is time-dependent\cite{budd}. Our purpose here is to
examine the general case with the magnetic translation concept as an
essential tool and to provide a rigorous demonstration of the
quantum adiabatic theorem involving infinitely degenerate energy
levels.

With the above physical picture, it is reasonable to resist any
temptation to believe that a wave packet initially centered at ${\bf
R}(0)$ will be displaced to be centered at point ${\bf R}(t)$ at
time $t$, almost as one should resist a temptation to believe that
the wave packet is not being displaced at all and it's just the
wavefunction that undergoes a time dependent local gauge
transformation that has no observable consequence. These two
contradicting scenarios have one thing in common: They can transform
an eigenfunction of the initial Hamiltonian to an instantaneous
eigenfunction of $H_L(t)$. Yet there are infinitely many
transformations that can achieve the same goal. Consider the local
gauge transformation scenario in which the initial eigenstate is not
displaced but is gauge transformed to be an instantaneous
eigenstate. One might perform a magnetic translation first before
doing the gauge transformation. In this paper we show that, under
this layer of local gauge transformation that depends on the value
of ${\bf R}(t)$, the motion of a wave packet is described by a
path-ordered magnetic translation corresponding to the displacement
$({\bf R}(t)- {\bf R}(0))/2$ and this path-ordered magnetic
translation is accompanied by a nonadiabatic operator that depends
on the rate of change of ${\bf R}(t)$. This nonadiabatic operator is
explicitly constructed for general parameter variation.

The Schr\"{o}dinger equation that corresponds to $H_L$ is
\begin{equation}
i{\hbar}\frac{\partial}{\partial t}\Psi_L({\bf x},t)=H_L\Psi_L({\bf
x},t).
\end{equation}
Now consider the gauge transformation
\begin{equation}
\Psi_L({\bf x},t)= \exp[-i\frac{q}{\hbar c}\chi({\bf x},t)]\Psi({\bf
x},t),
\end{equation}
where
\begin{equation}
\chi({\bf x},t)=
-\frac{B}{2}(R_2(t)-R_2(0))x_1+\frac{B}{2}(R_1(t)-R_1(0))x_2.
\end{equation}
Then $\Psi({\bf x},t)$ can be shown to satisfy the Schr\"{o}dinger
equation
\begin{equation}
i{\hbar}\frac{\partial}{\partial
t}\Psi({\bf x},t)=H\Psi({\bf x},t),
\end{equation}
where
\begin{equation}
H =\frac{1}{2m}\big[{\bf p}-\frac{q}{c}{\bf A}({\bf x})\big]^2
-\frac{q}{c}\frac{\partial}{\partial t}\chi({\bf x},t),
\end{equation}
\begin{equation}
{\bf A}({\bf x})= \frac{B}{2}{\bf e}_3\times\big[{\bf x}-{\bf
R}(0)\big].
\end{equation}
This particular gauge transformation, the one we mentioned earlier
that transforms any initial eigenfunction to an instantaneous
eigenfunction, induces a transformation between the time evolution
operators:
\begin{equation}
U_L(t,0)=\exp[-i\frac{q}{\hbar c}\chi({\bf x},t)]U(t, 0).
\end{equation}
Our goal is to find the factorization of $U(t, 0)$ corresponding to
$H$ into the product of three factors, each of them has a distinct
physical meaning. From such a factorization, the structure of
$U_L(t, 0)$ is then known.

Using the gauge transformed $H$, we want to study the Heisenberg
evolution of $\pi_\mu$ and $\eta_\mu$. We observe that it is the
behavior of $\pi_\mu(t)$ and $\eta_\mu(t)$ instead of that of
$p_\mu(t)$ and $x_\mu(t)$ that directly leads to the factorization
of $U(t, t_0)$. We have
\begin{equation}
\dot{\pi}_\mu =\omega\epsilon_{\mu\nu}\pi_\mu -
\frac{qB}{2c}\epsilon_{\mu\nu}{\dot R}_{\mu}(t), \ \ \
\dot{\eta}_\mu = -\frac{qB}{2c}\epsilon_{\mu\nu}{\dot R}_{\mu}(t),
\end{equation}
where $\omega=\frac{qB}{mc}$, or
\begin{equation}
\dot{\pi}= -i\omega\pi +i \frac{qB}{2c}{\dot R}(t), \ \ \
\dot{\eta}_\mu = -\frac{qB}{2c}\epsilon_{\mu\nu}{\dot R}_{\mu}(t),
\end{equation}
where
\begin{equation}
\pi =\pi_1+ i\pi_2,\ \ \ R(t)=R_1(t)+ iR_2(t).
\end{equation}
The solution to the Heisenberg equations can then be expressed as
\begin{equation}
\pi(t)=\pi(0)e^{-i\omega t}+i \frac{qB}{2c}e^{-i\omega
t}\int\limits_{0}^{t}e^{i\omega s}\frac{d}{ds} R(s)ds,
\end{equation}
\begin{equation}
\eta_{\mu}(t)=\eta_{\mu}(0)-
\frac{qB}{2c}\epsilon_{\mu\nu}(R_{\nu}(t)-R_{\nu}(0)).
\end{equation}
It is very clear that the homogeneous terms in the expressions for
$\pi(t)$ and $\eta_{\mu}(t)$ can be generated by the usual dynamical
operator $D(t)=\exp(-iH_L(0)t/\hbar)$. To produce the extra terms in
the expression for $\pi(t)$, and $\eta_{\mu}(t)$, respectively using
an operator $W(t)$, such that $D(t)W(t)$ recovers the whole
solution, it suffices for $W(t)$ to satisfy:
\begin{equation}
W^{\dagger}(t)\pi(0)W(t)=\pi(0)+
i\frac{qB}{2c}\int\limits_{0}^{t}e^{i\omega s}\frac{d}{ds} R(s)ds,
\end{equation}
\begin{equation}
W^{\dagger}(t)\eta_{\mu}(0)W(t)=\eta_{\mu}(0)-
\frac{qB}{2c}\epsilon_{\mu\nu}(R_{\nu}(t)-R_{\nu}(0)).
\end{equation}
In view of the commutation relations (1), which imply $[\pi,
\pi^{\dagger}]=2\hbar qB/c$, and from the formula
$\exp(-B)A\exp(B)=A+[A,B]$ with the condition that $[A, B]$ commutes
with $A$ and $B$, it is clear that $W(t)$ can be chosen to be the
product of two mutually commuting operators, generated by $(1,
\pi(0), \pi^\dagger(0))$ and $(1, \eta_{1}(0), \eta_{2}(0))$
respectively. Each of theses operators produces a translation for
either $\pi(0)$ or $\eta_{\mu}(0)$ while leaving the other
unchanged. Writing $W(t)$ as $W(t)=K(t)M(t)$, we can make the
following choice for $K(t)$ and $M(t)$,
\begin{equation}
K(t)=T\exp\bigg(i\frac{\pi^{\dagger}(0)}{4\hbar}\int\limits_{0}^{t}e^{i\omega
s}\frac{d}{ds}R(s)ds+
i\frac{\pi(0)}{4\hbar}\int\limits_{0}^{t}e^{-i\omega
s}\frac{d}{ds}R^{*}(s)ds\bigg),
\end{equation}
\begin{equation}
M(t)=P\exp\big(-i{\hbar}^{-1}\eta_{\mu}(0)\frac{R_{\mu}(t)-R_{\mu}(0)}{2}\big),
\end{equation}
where $T\exp$ stands for time-ordered exponential. It's different
from the direct exponential by a numerical phase factor only,
similar to the path-ordered exponential. Therefore it can be
directly checked that $D(t)K(t)M(t)$ recovers the solutions to the
Heisenberg equations.

To verify that $D(t)K(t)M(t)$ not only recovers the solutions to the
Heisenberg equations for $\pi$ and $\eta_{\mu}$, but in fact is the
time evolution operator corresponding to $H$, we now verify that it
satisfies the Schr\"{o}dinger equation. Note that $M(t)$ commutes
with both $D(t)$ and $K(t)$, so we have
\begin{equation}
i{\hbar}\frac{\partial}{\partial t}\big(D(t)K(t)M(t)\big)
=i{\hbar}\big[\frac{\partial}{\partial
t}\big(D(t)M(t)\big)\big]K(t)+i{\hbar}M(t)D(t)\frac{\partial}{\partial
t}K(t).
\end{equation}
It is straightforward that
\begin{equation}
i{\hbar}\big[\frac{\partial}{\partial t}\big(D(t)M(t)\big)\big]K(t)
= \big(H_{0}(0)+\eta_{\mu}\dot{R}_{\mu}(t)/2\big)D(t)K(t)M(t).
\end{equation}
To calculate $i{\hbar}M(t)D(t)\frac{\partial}{\partial t}K(t)$,
first observe that
\begin{equation}
D(t)\pi(0)D^{\dagger}(t)=D^{\dagger}(-t)\pi (0)D(-t)=\pi
(0)e^{i\omega t},
\end{equation}
\begin{equation}
D(t)\pi^{\dagger}(0)D^{\dagger}(t)=\big(D(t)\pi(0)D^{\dagger}(t)\big)^{\dagger}=\pi^{\dagger}(0)e^{-i\omega
t}.
\end{equation}
Therefore
\begin{eqnarray}
i{\hbar}M(t)D(t)\frac{\partial}{\partial t}K(t)&=&\big(-\pi^{\dagger}(0)\dot{R}(t)/4-\pi(0)\dot{R}^{*}(t)/4\big)D(t)K(t)M(t),\nonumber\\
      &=&-\frac{1}{2}\big(\pi_{1}(0)R_{1}(t)+\pi_{2}(0)R_{2}(t)\big)D(t)K(t)M(t).
\end{eqnarray}
Combining terms and from the definitions of $\pi_\mu$ and
$\eta_\mu$, we now have
\begin{eqnarray}
i{\hbar}\frac{\partial}{\partial
t}\big(D(t)K(t)M(t)\big)&=&\big(H_L(0)+\frac{qB}{2c}x_1\dot{R}_2-\frac{qB}{2c}x_2\dot{R}_1\big)D(t)K(t)M(t),\nonumber\\
                        &=&H\big(D(t)K(t)M(t)\big).
\end{eqnarray}
Therefore we conclude that the time evolution operator corresponding
to $H$ is
\begin{equation}
U(t, 0)=D(t)K(t)M(t)=M(t)D(t)K(t).
\end{equation}
And the time evolution operator corresponding to the original
Hamiltonian $H_L$ is
\begin{equation}
U_L(t,0)=\exp[-i\frac{q}{\hbar c}\chi({\bf x},{\bf
R}(t))]M(t)D(t)K(t)
\end{equation}
Noteworthy is the explicit construction of the operator $K(t)$ that
depends on the rate of change of the parameter ${\bf R}(t)$. It
approaches the identity operator in the adiabatic limit of ${\dot
R}_{\mu}(t)\rightarrow 0$, though the formula is valid for a general
variation of the parameter, not necessarily adiabatically. $M(t)$ on
the other hand is a geometric operator. It is determined by the path
$C$ traversed by ${\bf d}(t)=({\bf R}(t)-{\bf R}(0))/2$, and
recovers a geometric phase ($\beta(C(\bf d))$ in equation (3)) for a
closed path when acting on an eigenstate.

What about nonadiabatic transitions? For a general variation of the
parameter, we have to resort to the general expression for $K(t)$.
For a slow variation of the parameter ${\bf R}(t)$, say ${\bf
R}(t)={\bf R}(\epsilon t)$, where $T=\frac{1}{\epsilon}$ is the
duration of the adiabatic process, perturbative treatment is
possible. The oscillating kernels in the integrals in the expression
for $K(t)$ make sure that the transition probability is of the order
$\epsilon^2$ for the entire duration $T$ of the adiabatic process.
Because $\pi^{\dagger}(0)$ and $\pi(0)$ have the meaning of being
proportional to the creation and annihilation operators on the
energy eigenstates, the transition probability should be of the
order of $n^2\epsilon^2/\omega^2$, during the entire adiabatic
process $0\leq t\leq T$, where $n$ is the energy quantum number of a
Landau level, $E_{n}=\hbar\omega(n+1/2)$. Therefore nonadiabatic
transition rates are energy level dependent and increase as $n^2$.

In conclusion, our construction of the time evolution operator
$U_L(t,0)$ can be characterized as seeing the magnetic translation
of a charged particle accompanied by nonadiabatic corrections
through a gauge transformation. It provides an example of how the
quantum adiabatic theorem is realized when infinitely degenerate
energy levels are involved. Since the factorization is valid for a
general time variation, it can be employed to study the evolution of
all kinds of initial states of the system and their geometric
phases, not only adiabatic evolutions of initial eigenstates.

\end{document}